\documentclass[twocolumn]{aastex631}

\newcommand{\Msun}{\,{\rm M_\odot}}
\newcommand{\fedd}{\,{f_{\rm Edd}}}
\newcommand{\Mblack}{M_\bullet}

\begin{document}

\title{Accretion from Winds of Red Giant Branch Stars May Reveal the Supermassive Black Hole in Leo I}

\author[0000-0001-9879-7780]{Fabio Pacucci}
\affiliation{Center for Astrophysics $\vert$ Harvard \& Smithsonian, Cambridge, MA 02138, USA}
\affiliation{Black Hole Initiative, Harvard University, Cambridge, MA 02138, USA}

\author[0000-0003-4330-287X]{Abraham Loeb}
\affiliation{Center for Astrophysics $\vert$ Harvard \& Smithsonian, Cambridge, MA 02138, USA}
\affiliation{Black Hole Initiative, Harvard University, Cambridge, MA 02138, USA}



\begin{abstract}

A supermassive black hole (SMBH) of $\sim 3\times 10^6 \Msun$ was recently detected via dynamical measurements at the center of the dwarf galaxy Leo I. Standing $\sim 2$ orders of magnitude above standard scaling relations, this SMBH is hosted by a galaxy devoid of gas and with no significant star formation in the last $\sim 1$ Gyr. This detection can profoundly impact the formation models for black holes and their hosts. We propose that winds from a population of $\sim 100$ evolved stars within the Bondi radius of the SMBH produce a sizable accretion rate, with Eddington ratios between $9\times10^{-8}$ and $9\times10^{-7}$, depending on the value of the stellar mass loss. These rates are typical of SMBHs accreting in advection-dominated accretion flow (ADAF) mode. The predicted spectrum peaks in the microwaves at $\sim 0.1-1$ THz ($300-3000 \, \mathrm{\mu m}$) and exhibits significant variations at higher energies depending on the accretion rate. We predict a radio flux of $\sim 0.1$ mJy at $6$ GHz, mildly dependent on the accretion properties. Deep imaging with Chandra, VLA, and ALMA can confirm the presence of this SMBH and constrain its accretion flow.

\end{abstract}



\section{Introduction} 
\label{sec:intro}

The dwarf spheroidal (dSph) galaxy Leo I hosts, at its center, a supermassive black hole (SMBH) of $(3.3 \pm 2) \times 10^6 \Msun$, according to a recent study by \cite{Bustamante-Rosell_2021}. The presence of the SMBH was determined dynamically by considering the central kinematics and analyzing a steady increase in the velocity dispersion toward the center. The absence of a central black hole is excluded at $95\%$ significance in all the various dynamical models considered. In this Letter, we label this central SMBH Leo I*.

Leo I is a close-by dSph galaxy, only $\sim 255 \, \mathrm{kpc}$ away, and with a virial mass of $(7\pm 1)\times 10^8 \Msun$ \citep{Mateo_2008}. The discovery of an SMBH in such a small galaxy is remarkable. A black hole of $\sim 3 \times 10^6 \Msun$ in Leo I places the system $\sim 2$ orders of magnitude above the standard relations between black hole mass and host properties (e.g., \citealt{Ferrarese_2000, Gebhardt_2000, Kormendy_2013}). 

Such an overmassive black hole raises questions concerning the origin of this discrepancy.
Several studies suggested that the presence of a central massive black hole may be a consequence of the formation process of dSph galaxies (e.g., \citealt{Volonteri_Perna_2005, Kormendy_2013, Amaro-Seoane_2014, Silk_2017}), with many of which hosting an actively accreting black hole \citep{Pacucci_2021}. In particular, \cite{Amaro-Seoane_2014} argued that the presence of overmassive black holes in dSph galaxies can be explained by their formation being triggered by dynamical interactions with young stellar clusters, which would eventually sink the SMBH at the center of the stellar distribution. The presence of overly massive black holes in dSPh galaxies could thus be regarded as a signature and not an exception. 

In Leo I, the presence of an SMBH similar to Sgr A* \citep{EHT_2022} escaped detection until recent times; the extreme properties of its host can explain this. Leo I is almost devoid of gas \citep{Mateo_2008} — it is a fossil galaxy whose star formation came to an almost complete halt $\sim 1$ Gyr ago due to ram pressure stripping during its infall toward the Milky Way. The residual star formation surface density is estimated to be $\sim 10^{-10} \Msun \, \mathrm{yr^{-1} \, pc^{-2}}$ \citep{Ruiz-Lara_2021}. Recent studies (e.g., \citealt{Regan_2022}) have suggested that the presence of a central SMBH and several off-centered massive black holes in a fossil galaxy would be indicative of heavy seed formation channels in the early Universe (see, e.g., \citealt{Woods_2018, Inayoshi_2020}).

In this Letter, we investigate the possibility of detecting Leo I* electromagnetically. In \S \ref{sec:model}, we show that accretion from the winds of the red giant branch (RGB) stellar population in Leo I can produce a sizable accretion rate.
In \S \ref{sec:SED}, we describe the model for the spectral energy distribution and its limitations.
In \S \ref{sec:observations}, we discuss the resulting electromagnetic signature with predictions for the detection of Leo I* in several bands. Finally, in \S \ref{sec:conclusions}, we describe why we rule out the possibility of using gravitational lensing to reveal the presence of the SMBH. We conclude by proposing future observational campaigns directed at this remarkable dwarf galaxy.

\section{Accretion Model}
\label{sec:model}
The mass of Leo I* is estimated to be $\Mblack \approx 3.3 \times 10^6 \Msun$ \citep{Bustamante-Rosell_2021}, while the stellar velocity dispersion of the dwarf is $\sigma \sim 10 \, \mathrm{km \, s^{-1}}$ \citep{Mateo_2008}. The resulting Bondi radius of the SMBH is 
\begin{equation}
   R_B = \frac{2G\Mblack}{c^2_s} \approx 280 \, \mathrm{pc} \, ,
\end{equation}
under the assumption that the sound speed $c_s$ of virialized gas is equal to the velocity dispersion in the core of Leo I.

Remarkably, the core radius of Leo I is $R_{\rm core} = 245 \pm 35$ pc \citep{Mateo_2008}; the entire core of the dwarf is within the gravitational sphere of influence of Leo I*. We are unaware of any other galaxy whose core is gravitationally dominated by the central SMBH.

Located at a heliocentric distance of $255$ kpc \citep{Mateo_2008}, the Bondi radius subtends an angle of $226$ arcseconds. Within this angular distance, there are $106$ kinematically confirmed and $480$ photometrically confirmed RGB stars \citep{Mateo_2008}. From \cite{Meszaros_2009} and \cite{Mullan_2019} we derive that the average mass-loss rate on RGB stars is in the range $10^{-9} \Msun \, \mathrm{yr^{-1}}$ to $10^{-8} \Msun \, \mathrm{yr^{-1}}$. Note that this is at least $\sim 10^5$ times larger than the solar value of $2 \times 10^{-14} \Msun \, \mathrm{yr^{-1}}$ \citep{Wood_2005}.
Due to the low surface gravity of RGB stars, wind velocities at the tip of this stellar evolution phase are generally lower than $10 \, \mathrm{km\, s^{-1}}$ \citep{Yasuda_2019}, but the wind terminal velocity can be as high as $20 \, \mathrm{km\, s^{-1}}$ \citep{Mellah_2020}.

ADAF accretion models are characterized by strong winds. As a consequence, less than $\sim 1\%$ of the gas available at the Bondi radius is accreted onto the black hole, while most of it is lost \citep{YWB_2012, Yuan_2015}.
In particular, \cite{YWB_2012} argue that the accretion rate flowing in onto the black hole scales as $\dot{\Mblack}_{\rm in}(r) \propto r^{0.5}$. Using the radial distribution of RGB stars provided in \cite{Mateo_2008}, we calculate the mass fraction that effectively flows into the black hole. We divide the extent of the Bondi radius into 100 bins; for each bin at radial distance $r$, we compute the number of stars it contains from \cite{Mateo_2008} and propagate inward the accretion rate with $\dot{\Mblack}_{\rm in}(r) \propto r^{0.5}$. We find that $\sim 6\%$ of the gas available is actually accreted onto the black hole. This fraction is somewhat higher than the $<1\%$ mentioned above because the stars are distributed well within the Bondi radius \citep{Mateo_2008}.

Hence, we calculate the accretion rate $\dot{\Mblack}$ over the central SMBH to be in the range: $6 \times 10^{-9} < \dot{\Mblack} \, [\Msun \, \mathrm{yr^{-1}}] < 6 \times 10^{-8}$. Given an Eddington rate of $\dot{M}_{\rm Edd} = 7.3 \times 10^{-2} \Msun \, \mathrm{yr^{-1}}$ for a $3.3 \times 10^6 \Msun$ black hole, we infer a range of Eddington ratios, $\fedd \equiv \dot{\Mblack}/\dot{M}_{\rm Edd}$, of $9 \times 10^{-8} < \fedd < 9 \times 10^{-7}$.

\begin{figure*}
\includegraphics[angle=0,width=\textwidth]{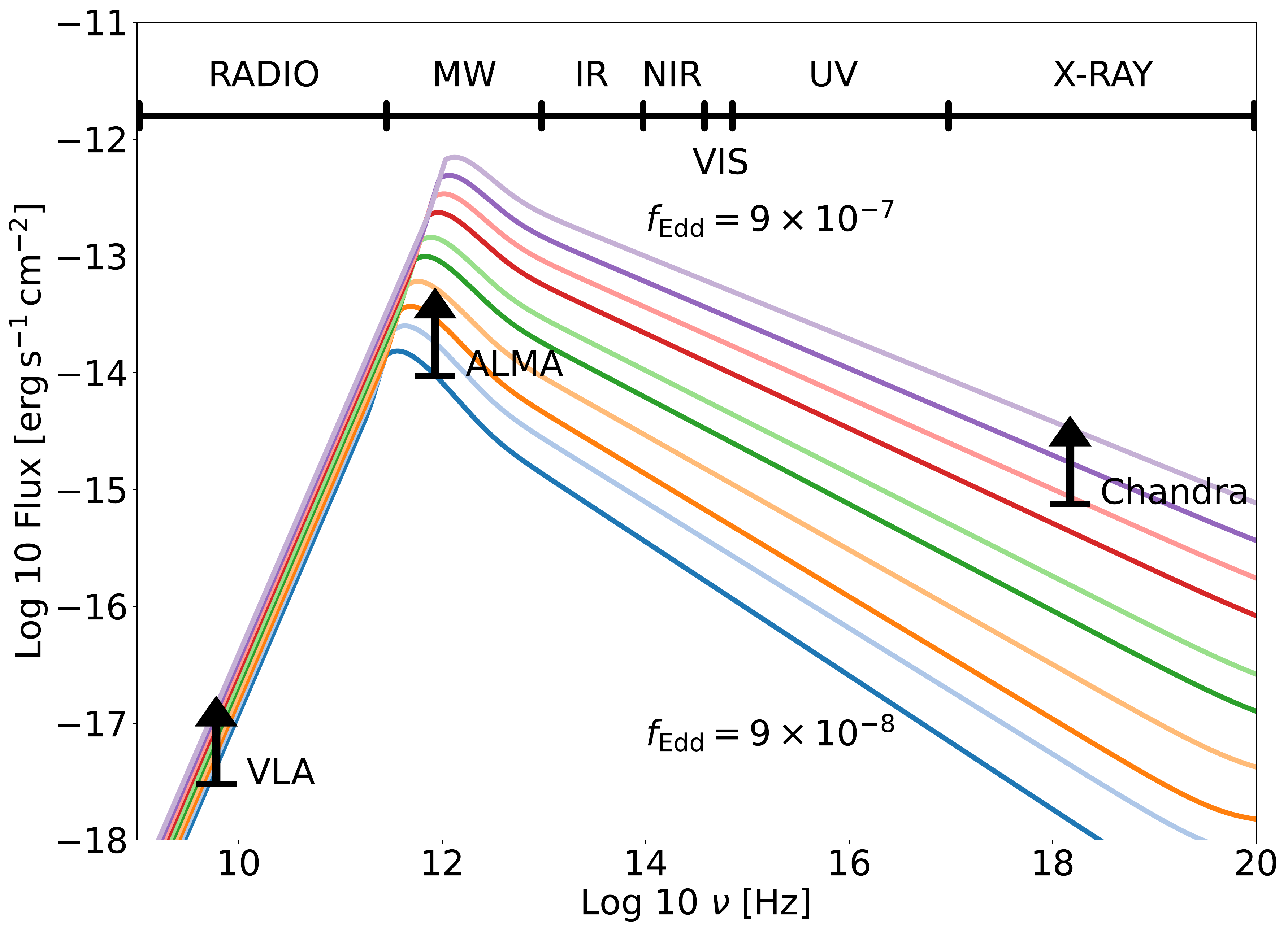}
\caption{Spectral energy distributions for the SMBH Leo I*, computed for ten values of the Eddington ratio, equally spaced logarithmically between $9 \times 10^{-8}$ and $9 \times 10^{-7}$. Continuum sensitivities for the VLA, ALMA, and Chandra are displayed. For the VLA, we consider its 6 GHz band and an observation time of $10$ minutes. For ALMA, we consider its $850$ GHz band (band 10) and an integration time of 1 minute. For Chandra, we assume photon energy of $6$ keV and an integration time of $35$ ks, obtaining $\sim 10$ photons.}
\label{fig:SED}
\end{figure*}

\section{Spectral Energy Distribution}
\label{sec:SED}
For $\fedd \ll 1$ we enter the domain of advection-dominated accretion flows or ADAF \citep{Narayan_1994, Narayan_1995, Abramowicz_1995, Narayan_2008, Yuan_Narayan_2014}. The vast majority of black holes in the local Universe accrete in ADAF mode, including the SMBH at the center of our Galaxy (e.g., \citealt{Yuan_2003}) and, possibly, intermediate-mass black holes wandering inside it \citep{Seepaul_2022}. Very low radiative efficiencies characterize these accretion flows. 

To simulate the spectral energy distribution (SED) radiated from the SMBH in Leo I  accreting in ADAF mode, we use an analytical model described in \cite{Pesce_2021}, which is based on the original 
work by \cite{Mahadevan_1997} with some modern update. The range of accretion rates of interest for our work has been widely explored by complex general-relativistic magnetohydrodynamic (GRMHD) simulations, characterized by accretion rates as low as $\dot{\Mblack} \sim 10^{-9} \dot{M}_{\rm Edd}$ \citep{Ryan_2017, Ressler_2017, Chael_2018, Chael_2019}.

We caution the reader of some limitations of our approach.
First, the X-ray emission from the system can be dominated by a jet rather than by the ADAF itself for extremely low accretion rates \citep{Yuan_Cui_2005, Yuan_2009}. The model described in \cite{Pesce_2021} does not include jets. 
Active galactic nuclei are generally divided into jetted and non-jetted \citep{Padovani_2016}; the differences between the two classes extend from the radio to the X-ray and $\gamma$-ray parts of the spectrum. If the SMBH in Leo I is characterized by a sizable jet, then its emission must be considered to improve our predictions for the SED.
Second, for extremely low accretion rates, corrections in the high-energy part of the SED are necessary and characteristic bumps due to bremsstrahlung radiation appear \citep{Yuan_Narayan_2014}. The model described in \citep{Pesce_2021} shows the appearance of such high-energy bumps for $\fedd \lesssim 10^{-8}$. 
In order to provide an SED that is tailored to the case of Leo I*, we need to: (i) estimate its actual Eddington ratio and (ii) assess the presence of a jet.
We were awarded Chandra (nr. 27480) and VLA (nr. 22B-297) Director's Discretionary Time (DDT) to provide an answer to these crucial unknowns.

\section{Electromagnetic Detection}
\label{sec:observations}

The SED predicted by our RGB-fueled accretion model is shown in Fig. \ref{fig:SED} for ten values of the Eddington ratio, bracketing the maximum and minimum values predicted in our framework.
Two features of these SEDs are noteworthy:
\begin{itemize}
    \item The peak emission falls in the microwave part of the electromagnetic spectrum, independently of the accretion rate. While the peak of the SED shifts to higher frequencies with increasing accretion rates, its values hover around $0.1-1$ THz ($300-3000 \, \mathrm{\mu m}$).
    \item The radio emission below $\sim 100$ GHz is not affected by the accretion rate, providing a robust prediction of the model. In contrast, the SED at frequencies higher than the peak depends significantly on the accretion rate. In particular, the X-ray emission at $2$ keV ($\approx 4.8 \times 10^{17}$ Hz) varies by more than three orders of magnitude within our range of expected accretion rates.
\end{itemize}

\begin{figure}
\includegraphics[angle=0,width=0.49\textwidth]{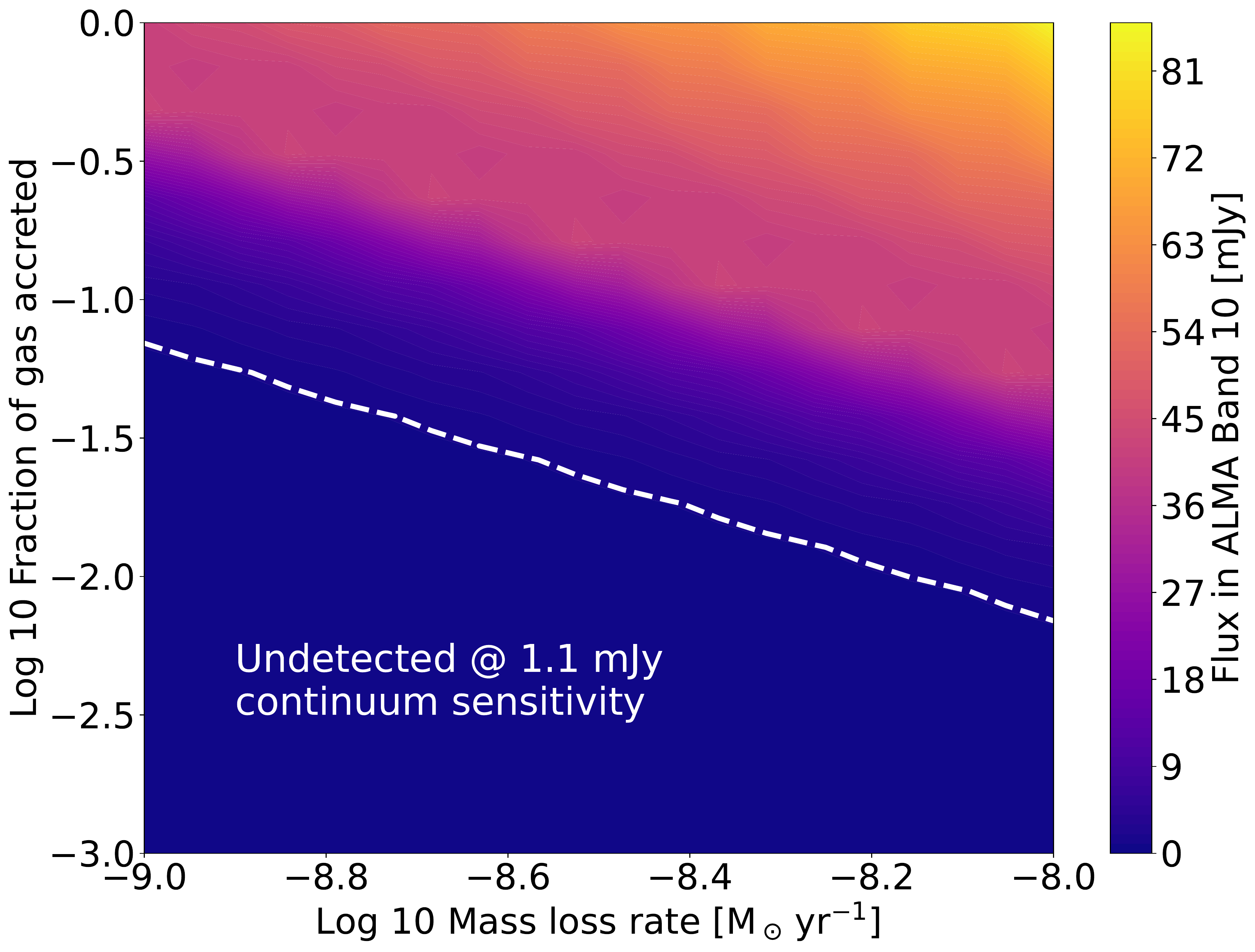}
\caption{Detectability analysis of the peak microwave emission of Leo I* by ALMA in band 10 ($\sim 850$ GHz, $\sim 1.1$ mJy continuum sensitivity). The parameter space includes mass-loss rates and the fraction of gas accreted from RGB stars within the Bondi radius of Leo I*. The source is undetected at $1.1 \, \mathrm{mJy}$ continuum sensitivity below the dashed line.}
\label{fig:contour}
\end{figure}

As the peak emission falls in the microwave regime, the Atacama Large Millimeter/submillimiter Array (ALMA) would provide an excellent opportunity to detect this source, although it would not likely provide a final word regarding its nature. In Fig. \ref{fig:contour}, we display a detectability analysis for the peak emission of Leo I* in the ALMA band 10 ($\sim 850$ GHz), the one that most closely traces the peak. We assume a continuum sensitivity of $1.1$ mJy with integration time as low as 1 minute, as reported in ALMA technical papers \citep{ALMA_2009}.
We perform the analysis as a function of two parameters: (i) the average mass-loss rate for RGB stars within the Bondi radius of Leo I*; (ii) the fraction of gas lost from these stars, which the SMBH eventually accretes. Note that our model assumes a value of $6\%$, due to strong winds characteristic of the ADAF model. However, even with fractions $<1\%$ \citep{YWB_2012}, ALMA would detect the SMBH with a very short integration time if the mass-loss rate is $>10^{-8} \, \mathrm{\Msun \, yr^{-1}}$.
In summary, an ALMA detection, although not conclusive to establish the nature of the source, would provide a simple soundness check for the SMBH hypothesis in Leo I.

The radio emission predicted in our model depends weakly on the accretion rate. For example, in the Very Large Array (VLA) $\sim 6$ GHz band, we predict a radio flux of $0.1 \, \mathrm{mJy}$, detectable with $\sim 10$ minutes of integration with the VLA in D configuration.
On the contrary, the X-ray flux depends strongly on the value of the accretion rate, and it may or may not be readily observable with current observatories. Clear X-ray detection of this source would provide a bedrock foundation to build the case of the existence of Leo I*. The measurement of the X-ray flux would also be fundamental to determining its accretion rate. 

An X-ray detection in the dynamically measured location of Leo I* would firmly indicate the presence of the SMBH — alternatives, in fact, are scarce. First, given the source density, a background AGN is highly unlikely. The alternative of a stellar origin to the X-ray emission is possible but also unlikely. Note that the ancient stellar populations of Leo I (and, in general, of dSph galaxies) make the presence of X-ray binaries, both low mass and high mass, unlikely. Additionally, low-mass X-ray binaries were studied in Leo I and no source identified as such is present within the uncertainty radius of Leo I* (see, e.g., \citealt{Orio_2010}). Moreover, in a dwarf spheroidal galaxy of the same mass as Leo I, namely the dSph Draco, the presence of X-ray sources was studied extensively \citep{Saeedi_2019}, leading to the discovery of only three symbiotic stars in the galaxy. This scarcity suggests that, although possible, the presence of stellar-type sources within the uncertainty region for Leo I* is improbable.

\section{Discussion and Conclusions}
\label{sec:conclusions}
This Letter is motivated by the discovery, by dynamical measurements, of an SMBH in the dSph galaxy Leo I \citep{Bustamante-Rosell_2021}. We showed that accretion from a small population of RGB stars inside the Bondi radius of the SMBH could produce an electromagnetic emission sufficient to make it detectable. We predict an SED that peaks in the microwave band and, at higher energy, depends strongly on the accretion rate. In contrast, the radio emission is mildly dependent on the specific accretion rate and provides a solid test for our model.

Gravitational lensing of a massive foreground object on background light sources was recently used to detect, for the first time, the presence of an isolated black hole in the Milky Way \citep{Lam_2022, Sahu_2022}.
A similar detection pathway is unlikely for Leo I*. First, the stars belonging to Leo I are too close to the central SMBH to produce a sizable Einstein radius. Cosmological sources are viable candidates to be lensed, but their surface density is too low. The Einstein radius is calculated from $D_{LS}$, $D_{L}$, and $D_{S}$, which are the various distances to and between the source (S) and the lens (L).
Assuming $D_{LS} \sim D_S \gg D_L$, we calculate an asymptotic Einstein radius of $0.2$ arcseconds. We use the Hubble Ultra Deep Field \citep{HUDF_2006} as a reference, as it contains a surface density of galaxies of $\sim 0.24 \, \mathrm{arcsec^{-2}}$. Within an Einstein radius of $0.2$ arcsec, we expect about $0.03$ galaxies. Increasing the limiting magnitude from 30 (the Ultra Deep Field) to $\sim 34$ (the JWST) will not improve the situation, as the expected number of galaxies plateaus. We conclude that it is implausible to detect the presence of Leo I* via gravitational lensing effects.

A careful search of the electromagnetic signature of Leo I* will likely be successful. The electromagnetic detection would represent a landmark in the study of black holes. This second-closest SMBH, after Sgr A*, would constitute a unique laboratory to study accretion at very low rates. The dynamics of an SMBH hosted by such a tiny galaxy would also be worth investigating. Finally, in the long term, the space-borne, next-generation Event Horizon Telescope \citep{Pesce_2021} might be able to image it directly.

\begin{acknowledgments}
We thank the referee and the scientific editor for their insightful comments on the paper.
F.P. acknowledges fruitful discussions with Nico Cappelluti, Andrea Dupree, Giacomo Fragione, and Genevieve Schroeder. F.P. also acknowledges support from a Clay Fellowship administered by the Smithsonian Astrophysical Observatory. This work was also supported by the Black Hole Initiative at Harvard University, which is funded by grants from the John Templeton Foundation and the Gordon and Betty Moore Foundation.
\end{acknowledgments}

\bibliography{ms}{}
\bibliographystyle{aasjournal}



\end{document}